%% file: ds19.tex
\documentclass[a4paper,11pt]{article}

\usepackage[latin1]{inputenc}
\usepackage[OT1]{fontenc}
\usepackage{amsmath}
\usepackage{amssymb}
\usepackage{amscd}
\usepackage{amsthm}
\usepackage[usenames,dvipsnames]{xcolor}
\usepackage{textcomp}
\usepackage{tabularx}
\usepackage{wrapfig}
\usepackage{array}  
\usepackage{tikz} 
\usepackage{mathrsfs}
\usetikzlibrary{snakes,shapes,backgrounds,arrows,matrix,petri} 
\usepackage{extarrows}
\usepackage{amsmath}\usepackage{stmaryrd} 
\usepackage{graphicx}

\def\B{\ensuremath{\mathbb{B}}}
\def\Bn{\ensuremath{\B^n}}
\def\N{\ensuremath{\mathbb{N}}}

\newtheorem{theorem}{Theorem}

\begin{document}

\renewcommand{\floatpagefraction}{.95}

\setlength{\parindent}{0pt}

\title{About block-parallel Boolean networks: a position paper
}
  
\author{Jacques Demongeot$^{1}$, Sylvain Sen{\'e}$^2$\\[2mm]
  {\small $^1~$Universit{\'e} Grenoble Alpes, AGEIS EA 7407, 38000, Grenoble, 
		France}\\
  {\small $^2~$Aix-Marseille Univ., Toulon Univ., CNRS, LIS UMR7020, Marseille, 
		France}
}

\date{}
  
\maketitle

\begin{abstract}
	In automata networks, it is well known that the way entities update 
	their states over time has a major impact on their dynamics. In 
	particular, depending on the chosen update schedule, the underlying 
	dynamical systems may exhibit more or less asymptotic dynamical 
	behaviours such as fixed points or limit cycles. Since such 
	mathematical models have been used in the framework of biological 
	networks modelling, the question of choosing appropriate update 
	schedules has arised soon. In this note, focusing on Boolean networks, 
	our aim is to emphasise that the adequate way of thinking regulations 
	and genetic expression over time is certainly not to consider a wall 
	segregating synchronicity from asynchronicity because they actually 
	complement rather well. In particular, we highlight that specific 
	update schedules, namely block-parallel update schedules, whose 
	intrinsic features are still not known from a theoretical point of view, 
	admit realistic and pertinent properties in the context of biological 
	modelling and deserve certainly more attention from the community.\\[2mm]
\end{abstract}

\section{Introduction}

Automata networks (ANs) are discrete dynamical systems, introduced in 1943 
by McCulloch and Pitts~\cite{McCulloch1943}, widely used to model 
genetic control networks and more generally biological networks since the 
end of the 1970's and the seminal works of 
Kauffman~\cite{Kauffman1969b,Kauffman1969,Kauffman1974} and 
Thomas~\cite{Thomas1973,Thomas1978,Thomas1981}. In this context of 
molecular systems biology, the ANs considered are finite, \emph{i.e.}, they 
are composed of a finite number of nodes (or automata) that interact with 
each other over a discrete time.

As soon as they have been used as models of genetic regulation networks, the 
way that nodes had to be updated over time was discussed. Indeed, when 
Kauffman introduced them at the end of the 1960's by scheduling nodes 
in parallel (certainly due to mathematical usabilities), biological arguments 
were highlighted notably by Thomas arguing that parallelism was not likely 
because of the impossibility for genetic expressions to occur 
simultaneously. Of course, the links between parallelism and simultaneity 
could be discussed further in the context of modelling but this is not the 
purpose of this note. So, other ways of updating ANs were introduced that 
are located at the other end of the spectrum, based on the concept of 
asynchronicity such as the asynchronous method of Thomas et 
al.~\cite{Remy2008,Thomas1973,Thomas1991}, and random sequential
ANs~\cite{Demongeot1987,Gershenson2003,Harvey1997,Savage2005}. This created 
a strong separation between the synchronous and sequential ways of 
considering ANs, to such an extent that the schedules mixing together 
synchronicity and asynchronicity were set aside for decades in the context 
of modelling. Fortunately, it was not the case in mathematics and computer 
science, as shown in the works of 
Robert~\cite{Robert1969,Robert1980,Robert1986} on block-sequential update 
schedules, that were used in~\cite{Mendoza1998} in the framework of the 
modelling of the floral morphogenesis of the plant \emph{Arabidopsis 
thaliana} and began to be studied in depth from the second half of the 
2000's~\cite{Aracena2009,Demongeot2008,Elena2009}. 

In this note, following a natural computing approach that consists in using 
the well known biologically-inspired computational model of ANs, and 
studying instances of them to model real biological phenomena, our wish is 
to highlight that the update schedules studied until now (all belonging to 
the family of block-sequential update schedules), even if they are 
interesting theoretically, are far from being sufficient to capture specific
biological intricacies. In particular, biological timers (classically 
called ``Zeitgebers'' in biology and medicine) and clocks that can be of 
genetic or physiological nature/origin~\cite{Hardin1990,Hanse1992} need 
other ways of thinking updatings to be modelled. Here, our very aim is to 
put the emphasis on block-parallel update schedules, introduced initially 
in~\cite{Sene2008} and never studied \emph{per se} until now, by giving 
insights essentially, and to highlight that they have interesting 
theoretical properties because of their intrinsic complexity, and pertinent 
features from a modelling point of view since they allow to model biological 
timers.

Section~\ref{sec:prelim} recalls the basics related to ANs, and some of the
seminal results obtained in the past. In Section~\ref{sec:bp}, 
block-parallel update schedules are presented together with some of their 
very basic properties. Through two examples coming from distinct areas of 
biology, Section~\ref{sec:bio} underlines the ability of block-parallel 
modes to model biological timers. Finally, open questions are given in 
Section~\ref{sec:oq}.

\section{Preliminary}
\label{sec:prelim}

This section presents the classical notions and definitions used and widely 
studied in the literature related to Boolean networks. Of course, 
all of these can be easily generalised for other alphabets and multi-valued 
automata networks. Moreover, for the sake of simplicity without losing 
generality, we reduce the set of Boolean networks considered to threshold 
Boolean networks.

\paragraph{Threshold Boolean networks}

Let $V = \{1, ..., n\}$ a set of nodes (often called automata in the 
literature). A \emph{configuration} on $V$ is a one-to-one function $x : V 
\to \B = \{0,1\}$ so that a Boolean value is associated with each node of 
$V$. In other terms, a configuration $x = (x_1, x_2, ..., x_n)$ is a Boolean 
vector of dimension $n$, where any of the $x_i \in \B$ is the 
\emph{state} of node $i$ in configuration $x$. The discrete evolution of 
the local state of node $i$ is defined as a \emph{local transition 
function}
\begin{equation*}
	f_i : \left\lbrace\begin{array}{cll}
		\Bn & \to & \B\\
		x & \mapsto & H(\sum_{j=1}^{n} w_{i,j} \cdot x_j(t) - 
		\theta_i)\\
	\end{array}\right.\text{,}
\end{equation*}
where $W$ is the \emph{interaction matrix}, \emph{i.e.}, the real-valued 
square matrix of order $n$ such that coefficient $w_{i,j}$ is the weight of 
the influence that node $j$ has on node $i$, $\Theta$ is the 
\emph{activation vector}, \emph{i.e.}, the real-valued vector of dimension 
$n$ such that $\theta_i$ is the activation threshold of node $i$, and $H$ 
is the Heaviside step function such that $H(x) = 0$ if $x < 0$, and $1$ 
otherwise. Consequently, $f = (f_i)_{i \in \{1,...,n\}}$ is such that 
\begin{equation*}
	f : \left\lbrace\begin{array}{cll}
		\Bn & \to & \Bn\\
		x & \mapsto & H(W \cdot x - \Theta)\\
	\end{array}\right.\text{,}
\end{equation*} 
and defines the threshold Boolean network. 

\paragraph{Interaction graphs}

\begin{figure}[t!]
	\centerline{
		\begin{minipage}{.45\textwidth}
			\centerline{$f:$ \quad 
			$W = \begin{pmatrix}
				0 & 0 & 1\\
				1 & 0 & 0\\
				0 & 1 & 0\\ 
			\end{pmatrix}$ \quad $\Theta = \begin{pmatrix}
				+\varepsilon\\
				+\varepsilon\\
				+\varepsilon\\
			\end{pmatrix}$}
		\end{minipage}\quad\vrule\quad
		\begin{minipage}{.45\textwidth}
			\centerline{\scalebox{1}{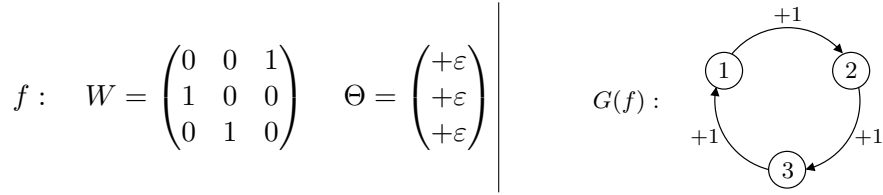}}
		\end{minipage}
	}
	\caption{(left) Definition of a network $f$ with its interaction matrix 
		$W$ and its threshold vector $\Theta$; (right) its associated 
		interaction graph $G(f)$.}
	\label{fig:cyclepos3}
\end{figure}

Let $f$ be a threshold Boolean network. From its interaction matrix $W$, we 
can derive the \emph{interaction graph} $G(f) = (V, I)$, with $I \subseteq 
V \times \mathbb{R} \times V$, where $(i,w_{j,i},j) \in I$ if $w_{j,i} \neq 
0$. When $w_{j,i} > 0$ (resp. $w_{j,i} < 0$, $w_{j,i} = 0$), node $i$ tends 
to activate (resp. tends to inhibit, has no influence on) node $j$. We 
generally speak  of positive or negative influences. 
Figure~\ref{fig:cyclepos3} depicts a threshold Boolean network composed
of $3$ nodes whose interaction graph is a cycle of size $3$. Notice that 
this network is more precisely a canonical \emph{positive 
cycle}~\cite{Demongeot2012,Noual2012,Sene2012,Thomas1981}, which means a 
cycle composed of an even number of negative influences (not to be confused 
with a \emph{negative cycle} which would be composed of an odd number of 
negative influences).

\paragraph{Update schedules}

In any configuration of a network, one or several punctual events may take 
place. Here, we consider events that consist in the update of at least 
one node state. Supposing that the network is currently in configuration 
$x \in \Bn$, node $i \in V$ is \emph{scheduled} if its state switches 
from $x_i$ to $f_i(x)$. Remark that, possibly, $f_i(x) = x_i$ so that 
the update of $i$ is not effective in $x$. In any case, this local event 
that we call a \emph{transition} yields a global network configuration 
change (possibly not effective) which is described by the 
\emph{$i$-update function} $F_i: \Bn \to \Bn$ such that
\begin{equation*}
  \label{EQ_iupdatefunction}
  \forall x \in \Bn,\ 
  	F_i(x) = (x_1,\ldots x_{i-1}, f_i(x), x_{i+1},\ldots, x_n)\text{.}
\end{equation*}
This transition is said to be atomic because it involves only one automaton. 
We also consider non-atomic transitions that correspond to the synchronous 
update of several nodes. In the general case, the \emph{$B$-update 
function}\footnote{$\forall i \in V,\, F_i$ 
obviously equals $F_{\{i\}}$.} $F_B : \Bn \to \Bn$ describes the 
network configuration change that results in the update of all the 
nodes of the subset (or block) $B$ of $V$ such that 
\begin{equation*}
  \label{EQ_Wupdatefunction}
  \forall x \in \Bn,\ \forall i \in V,\ F_B(x)_i = \begin{cases} 
		f_i(x) & \text{if } i \in B\\ 
		x_i & \text{otherwise}
  \end{cases}\text{.}
\end{equation*}
An \emph{update schedule} $\delta$ of a network whose set of nodes is 
$V$ is defined by an ordered (finite or infinite) sequence $(B_i)_{i \in 
\{0,\ldots,t-1\}}$ of $t$ non-empty subsets of nodes. We write $\delta = 
(B_i)_{i\in \{0,\ldots,t-1\}}$ or just $\delta = (B_0, B_1, \ldots, 
B_{t-1})$. Under an update schedule $\delta$, starting in configuration $x 
\in \Bn$, a network takes sequentially the configurations $x^0 = 
F_{B_0}(x)$, $x^1 = F_{B_1} \circ F_{B_0}(x)$, $\ldots$, $x^{t-1} = 
F_{B_{t-1}} \circ \ldots \circ F_{B_0}(x)$.

\emph{Periodic update schedules} of arbitrary period $p \in \N$
are infinite periodic sequences $(B_0, B_1, \ldots, B_{p-1}, B_0, B_1,
\ldots, B_{p-1},\ldots)$. For the sake of simplicity, they are rather 
defined by finite ordered lists $(B_i)_{i\in \N / p \N}$ of size $p$: 
$\delta = (B_0, B_1, \ldots, B_{p-1})$. When $\bigcup_{i=0}^{p-1} 
B_i = V$, such update schedules are called \emph{fair update schedules} 
and are strong ergodic update schedules, the latter being defined as: 
there exists $m \in \N$ such that every node is updated in the time 
interval $\llbracket k; k+m \rrbracket,\ \forall k \in \N$. \emph{Global 
transition functions} $F[\delta]: \Bn \to \Bn$ related to such periodic 
update schedules are defined as
\begin{equation*}
  \label{EQ-GTF}
  \forall x \in \Bn,\ F[\delta](x) = F_{B_{p-1}} \circ \ldots \circ 
  F_{B_1} \circ F_{B_0} (x)\text{.}
\end{equation*}
Such very update schedule has never been studied in depth, certainly 
because of their inherent generality and underlying complexity. Remark nevertheless that they have been mentioned in~\cite{Goles2012}.

Well known instances of periodic update schedules are 
\emph{block-sequential schedules}~\cite{Demongeot2008,Elena2009,Goles2010,Goles2012,Robert1986,Robert1995}.
Their particularity lies in that their periodic sequence of updates involves 
exactly once each automaton of the network. Formally, a block-sequential 
update schedule of a network of node set $V$ is an ordered partition $P$ of 
$V$. With our notations, it can be defined as a finite sequence $(B_i)_{i \in \N / p \N}$ such that $V = \bigsqcup_{{i \in \N / p \N}} B_i$. It can 
also be defined as a function $\delta : V \to \N / p \N$. The \emph{parallel 
update schedule} is the unique block-sequential update schedule of period 
$p = 1$ ($\forall i \in V,\ \delta(i) = 1$). It updates all the nodes of
the network at each time step, simultaneously. The $n!$ \emph{sequential 
update schedules}~\cite{Mortveit2001,Reidys2006} are block-sequential 
update schedules with period equal to the size of the network 
($p = n$). They update only one node of the network at a time 
($\forall i \in \N / p \N,\ |B_i| = 1$). For the sake of clarity, 
when a network is subjected to a block-sequential schedule $\delta 
= (B_i)_{i \in \N / p \N}$, the nodes inside a subset of $P$ are 
updated simultaneously and the subsets are iterated sequentially 
at each time step, from $B_0$ to $B_{p-1}$.

\begin{figure}[t!]
	\centerline{
		\begin{minipage}{.25\textwidth}
			\centerline{\scalebox{1}{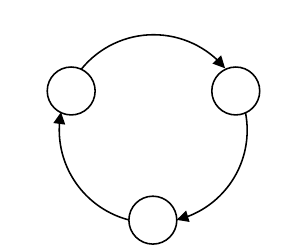}}
		\end{minipage}\quad\vrule\quad
		\begin{minipage}{.25\textwidth}
			\centerline{\scalebox{1}{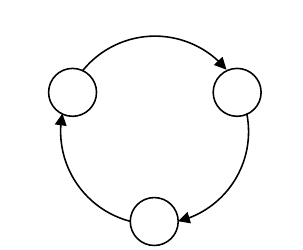}}
		\end{minipage}\quad\vrule\quad
		\begin{minipage}{.25\textwidth}
			\centerline{\scalebox{1}{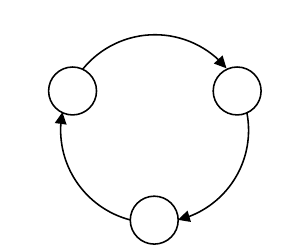}}
		\end{minipage}
	}
	\caption{Update graphs of the network $f$ defined in 
		Figure~\ref{fig:cyclepos3}: (left) Parallel update graph 
		$\mathscr{U}_{(\{1,2,3\})}(f)$; (centre) Block-sequential 
		update graph $\mathscr{U}_{(\{1,2\},\{3\})}(f)$; (right) 
		Sequential update graph $\mathscr{U}_{(\{1\},\{2\},\{3\})}(f)$.}
	\label{fig:cyclepos3_update}
\end{figure}

A non-classical way (but quite useful for proofs related to robustness 
of update schedules) to represent a block-sequential update schedule 
is the update graph introduced by Aracena et al. 
in~\cite{Aracena2013b,Aracena2013,Aracena2011,Aracena2009,Aracena2013c}. 
On the basis of $f$ a threshold Boolean network and the unlabelled 
version of its interaction graph $G'(f) = (V, I')$, where $I' = 
\{(i,j)\ |\ (i, w_{j,i}, j) \in I\}_{i,j \in V}$, a labelled graph 
$(G'(f), \text{lab})$ is defined, with $\text{lab} : I' \to 
\{<,\geq\}$. Given a block-sequential update schedule $\delta$ 
(seen as a function $\delta : V \to \N / p \N$), $U_\delta(f) = 
(G'(f), \text{lab})$ is the update graph associated with $\delta$ if:
\begin{equation*}
	\forall (i,j) \in I',\ \text{lab}(i,j) = \begin{cases}
		\geq & \text{if } \delta(i) \geq \delta(j)\\
		< & \text{if } \delta(i) < \delta(j)
	\end{cases}\text{.}
\end{equation*}
An illustration of the concept of update graphs is given in 
Figure~\ref{fig:cyclepos3_update}. From this update graph, Aracena et al. 
obtained the following very interesting result that gives strong insights on 
the role of block-sequential update schedules on the possible dynamical 
behaviours of a network.

\begin{theorem}[Aracena et al.~\cite{Aracena2009}]
	\label{thm:up_graph}
	Let $f$ be a Boolean network. Let $\delta_1$ and $\delta_2$ be two
	block-sequential update schedules. If $U_{\delta_1}(f) = 
	U_{\delta_2}(f)$ then $F[\delta_1] = F[\delta_2]$.
\end{theorem}

\paragraph{Transition graphs}

\begin{figure}[t!]
	\centerline{
		\begin{minipage}{.3\textwidth}
			\centerline{\scalebox{.9}{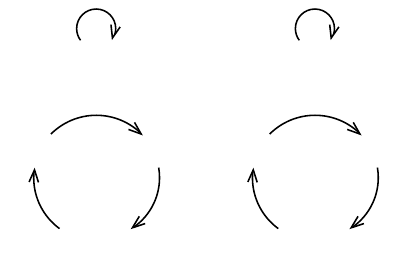}}
		\end{minipage}\quad\vrule\quad
		\begin{minipage}{.3\textwidth}
			\centerline{\scalebox{.9}{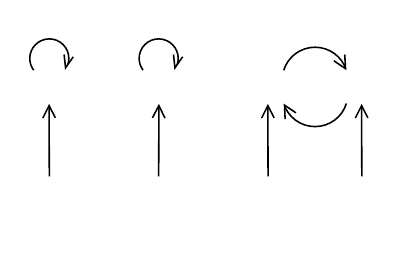}}
		\end{minipage}\quad\vrule\quad
		\begin{minipage}{.3\textwidth}
			\centerline{\scalebox{.9}{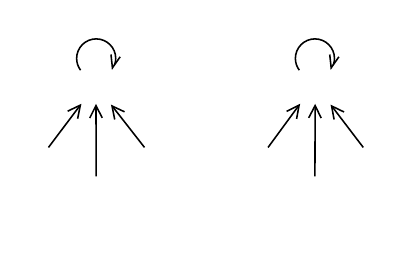}}
		\end{minipage}
	}
	\caption{Periodic dynamics of the network $f$ defined in 
		Figure~\ref{fig:cyclepos3}: (left) Parallel graph transition 
		$\mathscr{G}_{(\{1,2,3\})}(f)$; (centre) Block-sequential 
		graph transition $\mathscr{G}_{(\{1,2\},\{3\})}(f)$; (right) 
		Sequential graph transition 
		$\mathscr{G}_{(\{1\},\{2\},\{3\})}(f)$.}
	\label{fig:cyclepos3_trans}
\end{figure}

Given a network $f$ and a block-sequential update schedule $\delta$, 
the related \emph{transition graph} is $\mathscr{G}_\delta(f) = (\Bn, 
T_\delta)$, whose nodes are network configurations and arcs are network 
transitions such that $T_\delta \subseteq \Bn \times \Bn$. Formally, if 
$\delta = (B_i)_{i\in \N / p \N}$,  
\begin{equation*}
	\label{EQ-TGus}
	\mathscr{G}_\delta(f) = (\Bn, T_\delta) \text{ where } T_\delta = 
	\{(x, F[\delta](x))\ |\ x \in \Bn\}\text{.}
\end{equation*}
A transition graph $\mathscr{G}_\delta(f)$ represents the dynamics 
over time of network $f$ associated with update schedule $\delta$ (see 
Figure~\ref{fig:cyclepos3_trans}). From this graph, given a 
configuration $x \in \Bn$, the \emph{trajectory} of $x$ is the path 
(\emph{i.e.}, the transition sequence) that starts in $x$. Since the 
number of configurations is finite and $F[\delta]$ is a deterministic 
one-to-one function mapping $\Bn$ to itself, the trajectory of $x$ 
necessary ends up in a cycle of configurations, called an attractor. 
In this framework, an \emph{attractor} can be either a \emph{fixed point} of 
$F[\delta]$ (\emph{i.e.}, a stable configuration that repeats endlessly), 
or a \emph{limit cycle} of $F[\delta]$ (\emph{i.e.}, a sequence of recurrent 
configurations that repeats endlessly). Remark that, from a more applied 
point of view, a transition graph could be restricted to specific 
strict subsets of $V$ depending on the observability of the node 
states from which it could be relevant to compute and/or extract 
the related restricted attractors. Moreover, notice also that a transition 
graph can be extended by integrating all the intermediary visited
configurations resulted from the application of $F_{B_i}$, $\forall B_i \in 
(B_i)_{i \in \N / p \N}$. It is then called a \emph{complete transition 
graph}.

Let us recall now seminal and general theoretical results on 
\emph{(i)} the computable features and \emph{(ii)} the relations 
between the static (syntactic) and dynamical (semantic) properties 
of networks. 

\begin{theorem}[Goles and Martinez~\cite{Goles1990}]
	The computational model of Boolean networks is Turing-complete.
\end{theorem}

\begin{theorem}[Goles and Martinez~\cite{Goles1990}]
	\label{thm:pf}
	Let $f$ be a Boolean network, $\pi$ the parallel update schedule 
	and $\delta$ an arbitrary update schedule. If $x \in \Bn$ is a fixed 
	point of $F[\pi]$, then $x$ is a fixed point of $F[\delta]$.
\end{theorem}

Notice that the reciprocal of Theorem~\ref{thm:pf} is not true. A 
simple illustration of this consists in considering the positive 
cycle presented in Figure~\ref{fig:cyclepos3} that admits two fixed points, 
$000$ and $111$, when scheduled block-sequentially, and that admits the four 
following fixed points $000$, $010$, $101$, and $111$ when scheduled according to $\delta = (\{2,3\},\{1,3\},\{1,2\})$.

\begin{theorem}[Robert~\cite{Robert1986,Robert1995}]
	Let $f$ be a Boolean network associated with a directed 
	acyclic interaction graph. Whatever the update schedule $\delta$, 
	$F[\delta]$ admits a unique attractor that is a stable configuration.
\end{theorem}

The following theorem gives a strong global relation between interaction 
graphs and the existence of multi-stationarity of the underlying dynamical 
systems. It originally comes from a conjecture of Thomas presented 
in~\cite{Thomas1981}, was proven in the context of block-sequential 
update schedules by Aracena~\cite{Aracena2004} and then in the context of 
Thomas' asynchronous representation in~\cite{Remy2008b,Richard2007} before 
it was proven generally for any kind of update schedule by Noual and Sen{\'e} in~\cite{Noual2012,Sene2012}.

\begin{theorem}
	Let $f$ be a Boolean network and $G(f)$ its associated interaction 
	graph. Whatever the update schedule $\delta$, if $F[\delta]$ admits 
	several stable configurations then $G(f)$ contains a positive cycle.
\end{theorem}

\section{Block-parallel threshold Boolean networks}
\label{sec:bp}

In the previous section, we have emphasised that fair update schedules 
are very interesting from a theoretical point of view. Indeed, given 
a network $f$ and its interaction graph $G(f) = (V, I)$, they can make 
emerge peculiar asymptotic dynamics, such as stable configurations 
that are not fixed points of the parallel global transition function 
$F[(V)]$. However, on the one hand, such an intrinsic mathematical richness 
seems to be too much important from a more applied point of view directed 
toward the qualitative modelling of genetic regulation networks. Conversely, 
on the other hand, block-sequential update schedules are not sufficiently 
rich to model specific observed biological abilities. In particular, there 
exist some examples of genetic regulation networks in which a specific 
subnetwork plays a role of ``Zeitgeber'' (\emph{i.e.}, a timer) having its 
own clock. For instance, in Drosophila, this subnetwork consists in a small 
set of genes like TIME and PER~\cite{Goldbeter1995,Hardin1990,Sehgal1994} 
exhibiting as its unique attractor a limit cycle with its proper free-run 
period. This subnetwork influences other groups of genes, each of them 
having their own biological functionalities. Such a time command cannot be 
modelled by the classically studied block-sequential update schedules, 
simply because they prevent from having update repetitions of nodes in a 
same period of updates. Nevertheless, it can be modelled by another family 
of fair update schedules that we call block-parallel update schedules which 
are the topic of this note.

\begin{figure}[t!]
	\centerline{\scalebox{1}{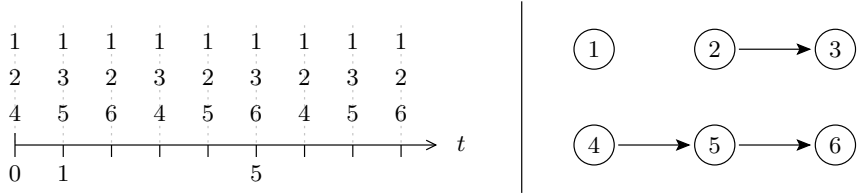}}
	\caption{(left) Time diagram of the updates and (right) partial 
		order graph of the anteriorities defined from the 
		block-parallel update schedule $\delta = \{(1),(2,3),(4,5,6)\}$.}
	\label{fig:anterioritiesgraph}
\end{figure}

Where a network $f$ scheduled block-sequentially evolves so that 
the nodes are updated simultaneously inside a block and the blocks 
themselves are iterated sequentially, a network scheduled 
block-parallelly evolves conversely. The blocks are iterated simultaneously 
and the nodes inside them are updated sequentially. In other terms, where a 
block-sequential update schedule is defined by an ordered partition of $V$ 
(\emph{i.e.}, a finite sequence of non-empty and disjoint subsets of $V$ 
recovering $V$), a block-parallel update schedule is defined by a set of 
non-empty and disjoint finite sub-sequences $S_i$ of $V$ whose union of 
elements recovers $V$. Thus, a block-parallel update schedule 
can be formally defined as $\delta = \{S_i\}_{i \in \{0, \ldots, 
|\delta|-1\}}$, such that $1 \leq |\delta| \leq |V|$. Remark that if 
$|\delta| = 1$ (resp. $|\delta| = |V|$), $\delta$ is a sequential (resp. the 
parallel) update schedule. Furthermore, notice that this definition 
satisfies the assumptions of a fair update schedule. Indeed, there exists a 
rewriting of any block-parallel update schedule into a finite sequence of 
subsets of $V$, complying with the dynamical properties of the underlying 
network. For instance, consider an arbitrary network of size $6$ with 
$V = \{1,...,6\}$ and the following block-parallel update schedule: 
$\{(1), (2,3), (4,5,6)\}$. The updatings along time of node states 
follow Figure~\ref{fig:anterioritiesgraph}(left), which corresponds 
exactly to the following finite sequence of subsets of $V$: 
\begin{equation*}
(\{1,2,4\},\{1,3,5\},\{1,2,6\},\{1,3,4\},\{1,2,5\},\{1,3,6\})\text{.}
\end{equation*}
A block-parallel update schedule can also be represented by a graph 
$\delta \equiv (V, A)$ that is simply the partial order graph 
associated with the partial order of the anteriorities defined by 
$\delta$ (see Figure~\ref{fig:anterioritiesgraph}(right)). 

Remark that block-parallel and block-sequential update schedules are 
distinct families of fair update schedules having a non-empty intersection 
made only of the block-sequential update schedules whose subsets have the 
same cardinality. Moreover, the union of these two families does not coincide 
with fair update schedules. Consider for instance a network $f$ of $3$ 
elements and $\delta = \{(0,1), (1,2), (0,1,2)\}$ that is not a 
block-parallel one since the sub-sequences are not disjoint. Its rewriting 
as a sequence of subsets is $(\{0,1\},\{1,2\},\{0,1,2\},\{0,1,2\},\{0,1\},
\{1,2\})$ which does not correspond to a block-sequential update schedule. 
However, it respects the definition of a fair update schedule. 

\section{Applications to biology}
\label{sec:bio}

Now the block-parallel update schedules have been introduced, let us 
highlight their relevance in the qualitative modelling of biological 
regulation networks. We present two applications, the first one in genetics 
and the second one in physiology. 

\subsection{Genetic control of plant growth}

The first example of a timer is the biological clock ruling the plant 
growth. A very schematic view of the functioning of the genetic control of 
the plant growth is to consider two components evolving independently. The 
first one can be modelled as a sub-network made of three genes that 
correspond to the localised expressions of the protein auxin (one of the 
plant growth regulators). The first gene AUX$a$ corresponds to the 
apical localisation, the others correspond to the axillary bud 
localisations, AUX$\ell$ for the left bud and AUX$r$ for the right bud. The 
second one can be modelled as a sub-network localised in the cotyledon and 
composed of two genes: CCA (Circadian Clock Associated gene) and TOC (Timing 
Of CAB expression gene).

\begin{figure}[t!]
	\centerline{
		\begin{minipage}{.55\textwidth}
			\centerline{\scriptsize $f:$\quad
			$W = \begin{pmatrix}
				1 & -2 & -2 & -2 & 0\\
				-2 & 1 & -2 & -2 & 0\\
				-2 & -2 & 1 & -2 & 0\\
				0 & 0 & 0 & 1 & -2\\
				0 & 0 & 0 & 1 & 0 
			\end{pmatrix}$ \quad $\Theta = \begin{pmatrix}
				-\varepsilon\\
				-\varepsilon\\
				-\varepsilon\\
				-\varepsilon\\
				\varepsilon
			\end{pmatrix}$}
		\end{minipage}\quad\vrule\quad
		\begin{minipage}{.425\textwidth}
			\centerline{\scalebox{1}{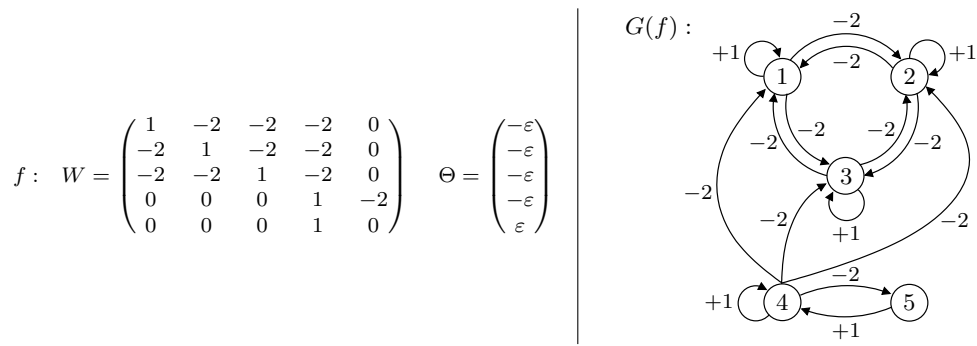}}
		\end{minipage}
	}
	\caption{(left) Definition of the genetic network 
		controlling the plant growth $f$ with its interaction 
		matrix $W$ and its threshold vector $\Theta$; (right) its 
		associated interaction graph $G(f)$.}
	\label{fig:netplant}
\end{figure}

The dynamics of the plant growth is governed by the threshold Boolean 
network $f$ where AUX$a \equiv 1$, AUX$\ell \equiv 2$, AUX$r \equiv 3$, CCA 
$\equiv 4$ and TOC $\equiv 5$, defined in Figure~\ref{fig:netplant} which is 
derived from~\cite{Bendix2015,Thellier2004}. Now, the idea is to focus 
only on the most realistic initial conditions for computing the dynamical 
behaviour. These conditions have to integrate the following biological 
observations: 
\begin{itemize}
\item In the first component, the apical auxin is the first expressed during 
	the plant growth. The other left and right bud auxins start at state $0$ 
	because the axillary buds are still not existing. 
\item In the second component, CCA is induced by photosynthesis (it starts 
	at state $1$) and activates secondarily TOC~\cite{Bendix2015}.
\end{itemize}
As a consequence, the most realistic initial configuration is $100\,10$. Let 
us consider now the update schedule. First of all, concerning the timer 
component, whatever the local update schedule, its dynamics remains the 
same. So, for the sake of simplicity, we have chosen that CCA and TOC evolve 
simultaneously. Now, concerning the auxin component, the messages 
transmitted from AUX$a$ to AUX$\ell$ corresponds to the diffusion along 
the stem of the auxin expressed by the gene AUX$a$. The same process can be 
observed between AUX$\ell$ and AUX$r$. As a consequence, a natural local 
update schedule for this component is the sequential one $(\{\text{AUX}a\},
\{\text{AUX}\ell\},\{\text{AUX}r\})$. To sum up, a realistic global update 
schedule of $f$ is the following block-parallel one:
\begin{equation*}
	\delta = \{(1,2,3), (4), (5)\}\ \iff\ \delta = 
	(\{1,4,5\}, \{2,4,5\}, \{3,4,5\})\text{,}
\end{equation*}
whose partial order graph is\\[2mm]
\centerline{\scalebox{.9}{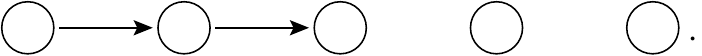}}\smallskip

\begin{figure}[t!]
	\centerline{
		\hspace*{2mm}\begin{minipage}{.45\textwidth}
			\begin{multline*}
			100\,10 \textcolor{gray}{\to 000\,11 \to 000\,01}\\ 
			\to \mathbf{010\,00} \textcolor{gray}{\to \mathbf{010\,10} \to 
				\mathbf{010\,11}}\\
			\to \mathbf{000\,01} \textcolor{gray}{\to \mathbf{100\,00} \to 
				\mathbf{100\,10}}\\
			\to \mathbf{100\,11} \textcolor{gray}{\to \mathbf{000\,01} \to 
				\mathbf{001\,00}}\\
			\to \mathbf{001\,10} \textcolor{gray}{\to \mathbf{001\,11} \to 
				\mathbf{000\,01}}\\
			\to 010\,00
			\end{multline*}
		\end{minipage}\qquad
		\begin{minipage}{.55\textwidth}
			\centerline{\includegraphics[scale=.4]{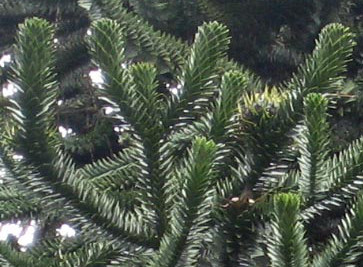}}
		\end{minipage}
	}
	\caption{(left) Trajectory of configuration $100\,10$ of network
		modelling the genetic control of plant growth governed by 
		the biological timer made of CCA and TOC; (right) Example 
		of the growth of \emph{Araucaria araucana} made by the 
		succession the triplets corresponding to an apical and two lateral 
		stems.}
	\label{fig:araucaria}	
\end{figure}

Exploiting all the knowledge given above about the networks, its interaction 
graph and block-parallel update schedule, the trajectory of the initial 
configuration is given in Figure~\ref{fig:araucaria}(left). It leads to a 
limit cycle of period $4$ (resp. of period $12$ if we consider it in the 
complete transition graph as it is the case in Figure~\ref{fig:araucaria}
(left)) starting in $01\,000$, whose internal structure made of every 
intermediary configuration comes from the combination of the three fixed 
points of the auxin component and of the limit cycle of period $4$ of the 
timer component. Notice also that the sequential growth of the three parts 
of the plant is correctly induced, in a uniform way. This regular scheme of 
growth could correspond to the quasi-perfect growth of \emph{Araucaria 
araucana} which consists in the succession of triplets made of a central and 
two lateral stems, as illustrated in Figure~\ref{fig:araucaria}(right).

\subsection{Cardio-respiratory regulation}


The second example of a timer is the biological clock ruling the 
cardio-respiratory regulation. As for the plant growth, the functioning of 
the physiological process of this regulation can be described by two 
components. The first one corresponds to the central vegetative system and 
plays the role of a timer. It comprises inspiratory ($I$) and expiratory 
($E$) neurons. The second component is peripheral. It comprises the 
sino-atrial node ruling the heart activity ($S$) and the baroreceptor 
located at the exit the left ventricle ($B$). 

\begin{figure}[t!]
	\centerline{
		\begin{minipage}{.45\textwidth}
			\centerline{\scriptsize
				$g:$\quad$W=\begin{pmatrix}
					0 & 2 & 0 & -1\\
					-2 & 1 & 1 & 0\\
					0 & -1 & 0 & 1\\
					1 & 0 & -1 & 2\\
				\end{pmatrix}$\quad$\Theta = \begin{pmatrix}
					\varepsilon\\
					-\varepsilon\\
					\varepsilon\\
					\varepsilon\\
				\end{pmatrix}$
			}
		\end{minipage}\quad\vrule\quad{}
		\begin{minipage}{.335\textwidth}
			\centerline{\scalebox{1}{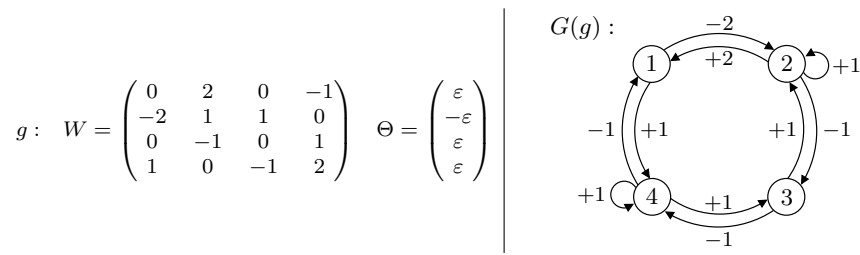}}
		\end{minipage}
	}
	\caption{(left) Definition of the physiologic network $g$
		controlling the cardio-respiratory regulation with its 
		interaction matrix $W$ and its threshold vector 
		$\Theta$; (right) its associated interaction graph 
		$G(g)$.}
	\label{fig:cardionet}
\end{figure}

The dynamics of the cardio-respiratory regulation is governed by the network 
$g$ where $E \equiv 1$, $I \equiv 2$, $B \equiv 3$, and $S \equiv 4$, 
defined in Figure~\ref{fig:cardionet} and derived 
from~\cite{Beauchaine2001,Demongeot2018,Dergacheva2010,Moraes2014}. 
Here also, let us consider the realistic initial condition $00\,00$ that 
corresponds to the end of an expiration just before the activity start of 
inspiratory neurons, and to low levels of the sino-atrial and baroreceptor 
activities. 

The local updating schedule of the timer is similar to that of a plant 
growth. That of the second component comes from the information transmission 
from $S$ to $B$ thanks to the blood flow. It is consequently sequential and 
defined by $(\{S\}, \{B\})$. To sum up, the realistic global update schedule 
of $g$ is the following block-parallel one:
\begin{equation*}
	\delta = \{(1), (2), (4, 3)\}\ \iff\ \delta = (\{1,2,4\},	
		\{1,2,3\})\text{.} 
\end{equation*}

The trajectory of initial realistic configuration $00\,00$ is
\begin{equation*}
	00\,00 \textcolor{gray}{\to 01\,00}
	\to 11\,00 \textcolor{gray}{\to 10\,01}\\
	\to \mathbf{00\,11} \textcolor{gray}{\to \mathbf{01\,11}}
	\to \mathbf{11\,01} \textcolor{gray}{\to \mathbf{10\,01}}\\
	\to 00\,11\text{.}
\end{equation*}
It suggests that the cardio-respiratory network presents a limit-cycle of 
period $2$ (resp. of period $4$ if we consider it in the complete transition 
graph as it is the case above with black and gray bold configurations) 
decomposed into two phases as in the real biological functioning. It is 
composed of the succession of the activity periods of the two families of 
central neurons ($E$ and $I$) that are both inactive and then active, and 
that can hence trigger the two intermediary inspiratory and expiratory 
phases. For the peripheral activity, we can observe a constant sino-atrial 
activity inducing an important cardiac activity at the end of the expiration 
and during the inspiration that is reduced after the inspiration and during 
the expiration. Such an observation could refer to the idea of a qualitative 
model for the well known phenomenon called cardiac sinusal arrhythmia.

A relevant characteristics of the network $g$ is that the reduction of the 
influence of the sino-atrial node on itself, by changing $w_{4,4} = 2$ to 
$w_{4,4} = 1$ suffices to retrieve the normal cardiac functioning with no 
sinus arrhythmia. Indeed, in this case, the trajectory of $00\,00$ becomes:
\begin{equation*}
	00\,00 \textcolor{gray}{\to 01\,00}
	\to \mathbf{11\,00} \textcolor{gray}{\to \mathbf{10\,01}}\\
	\to \mathbf{00\,11} \textcolor{gray}{\to \mathbf{01\,10}}
	\to 11\,00\text{.}
\end{equation*}
However, this trajectory highlights a phase shifting of the cardiac activity 
with respect to the firing of respiratory neurons of the vegetative system, 
which does not comply with the biological assumptions. As a consequence, a 
refining of the model seems to be required, which paves the way for further 
modelling studies. 

\section{Conclusion and open problems/questions}
\label{sec:oq}

This position paper aimed at putting the focus on block-parallel schedules 
and highlighting especially their ability to model autonomous timers that 
govern the dynamics of impacted sub-networks, for underlining their 
pertinence from a biological modelling point of view with respect to 
``Zeitgebers'' (timers). Beyond the modelling interest, these modes seem 
also to have very interesting theoretical properties that deserve without 
any doubt to be studied in depth from mathematical and theoretical computer 
science points of view, as it has been the case for block-sequential update 
schedules. Eventually, in order to illustrate our purpose, let us present 
open problems/questions which, if addressed, could lead to relevant advances 
in both theoretical and applied frameworks:
\begin{itemize}
\item In~\cite{Goles2010}, the authors showed that any block-sequential 
	isolated positive or negative cycle can be simulated by a smaller 
	parallel cycle of same sign. So, a natural question is to understand if 
	a similar property holds for block-parallel and another schedule that 
	the parallel one. Indeed, it is easy to see that the repetition of at 
	least one node updating is a lock here and that consequently, generally 
	speaking, a block-parallel cycle cannot be simulated by a smaller 
	parallel cycle. As a consequence, the open question is: as the parallel 
	is the canonical block-sequential update schedule for Boolean cycles, 
	does there exist a canonical block-parallel update schedule?
\item In~\cite{Demongeot2012}, by basing themselves on the result 
	of~\cite{Goles2010}, the authors proved all the combinatorial 
	transient and asymptotic properties of parallel Boolean cycles. If a 
	positive answer is found to the previous open question leading to the 
	definition of a canonical block-parallel update schedules, it will be 
	crucial to lead a similar combinatorial study to obtain a perfect 
	knowledge of cycles in this context since the latter remain important 
	in this framework in terms of dynamical complexity.
\item As presented in the note, 
	in~\cite{Aracena2013,Aracena2011,Aracena2009}, the authors introduced 
	block-sequential update graphs. This graph is of real interest since it 
	allows to know efficiently if a network associated with two 
	block-sequential update schedules admits an equivalent dynamics (see 
	Theorem~\ref{thm:up_graph}). In order to prove similar equivalence 
	properties in the context of block-parallel update schedules, it seems 
	that it would be of great interest to develop the concept of 
	block-parallel update graphs. Can such a concept be found? If yes, 
	given a network $f$ and a block-parallel update schedule $\delta$, is 
	it as easy to define/construct as for block-sequential update 
	schedules?
\item Concerning the attractors of block-parallel models with timers of real 
	biological networks, are the \emph{complete limit cycles} (\emph{i.e.}, 
	the limit cycles including intermediary configurations) always made of 
	an alternation of the attractors projected on the nodes of the 
	peripheral sub-networks provoked by the period of the limit cycle(s) of 
	the timer(s)? Even if it seems to be efficient, does nature tend to 
	classically use such timers/processes to make biological networks able 
	to oscillate between attraction basins and thus create functional 
	rhythms? To answer to these questions, a statistical study in depth of 
	well known networks modelling biological rhythms will be necessary. 
\end{itemize}

\paragraph{Acknowledgements} The present work has been partially supported 
by the ANR-13-TECS-0011 project ``e-swallhome'' (JD), and by the 
ANR-18-CE40-0002 ``FANs'', the PACA-15-APEX-01134 ``FRI'' and the ECOS C16E01 
projects (SS).

\bibliographystyle{alpha}
\bibliography{ds19.bib}

\end{document}

%% file: cyclepos3.pdf_t
\begin{picture}(0,0)%
\includegraphics{cyclepos3.pdf}%
\end{picture}%
\setlength{\unitlength}{3947sp}%
\begingroup\makeatletter\ifx\SetFigFont\undefined%
\gdef\SetFigFont#1#2#3#4#5{%
  \reset@font\fontsize{#1}{#2pt}%
  \fontfamily{#3}\fontseries{#4}\fontshape{#5}%
  \selectfont}%
\fi\endgroup%
\begin{picture}(1925,1190)(-627,-338)
\put(603,734){\makebox(0,0)[b]{\smash{{\SetFigFont{8}{9.6}{\familydefault}{\mddefault}{\updefault}{\color[rgb]{0,0,0}$+1$}%
}}}}
\put(1018,-30){\makebox(0,0)[lb]{\smash{{\SetFigFont{8}{9.6}{\familydefault}{\mddefault}{\updefault}{\color[rgb]{0,0,0}$+1$}%
}}}}
\put(603,-246){\makebox(0,0)[b]{\smash{{\SetFigFont{9}{10.8}{\familydefault}{\mddefault}{\updefault}{\color[rgb]{0,0,0}$3$}%
}}}}
\put(206,377){\makebox(0,0)[b]{\smash{{\SetFigFont{9}{10.8}{\familydefault}{\mddefault}{\updefault}{\color[rgb]{0,0,0}$1$}%
}}}}
\put(997,377){\makebox(0,0)[b]{\smash{{\SetFigFont{9}{10.8}{\familydefault}{\mddefault}{\updefault}{\color[rgb]{0,0,0}$2$}%
}}}}
\put(-194,191){\makebox(0,0)[rb]{\smash{{\SetFigFont{9}{10.8}{\familydefault}{\mddefault}{\updefault}{\color[rgb]{0,0,0}$G(f):$}%
}}}}
\put(178,-28){\makebox(0,0)[rb]{\smash{{\SetFigFont{8}{9.6}{\familydefault}{\mddefault}{\updefault}{\color[rgb]{0,0,0}$+1$}%
}}}}
\end{picture}%

%% file: cyclepos3uppar.pdf_t
\begin{picture}(0,0)%
\includegraphics{cyclepos3uppar.pdf}%
\end{picture}%
\setlength{\unitlength}{3947sp}%
\begingroup\makeatletter\ifx\SetFigFont\undefined%
\gdef\SetFigFont#1#2#3#4#5{%
  \reset@font\fontsize{#1}{#2pt}%
  \fontfamily{#3}\fontseries{#4}\fontshape{#5}%
  \selectfont}%
\fi\endgroup%
\begin{picture}(1433,1191)(-135,-339)
\put(1018,-30){\makebox(0,0)[lb]{\smash{{\SetFigFont{8}{9.6}{\familydefault}{\mddefault}{\updefault}{\color[rgb]{0,0,0}$\geq$}%
}}}}
\put(178,-28){\makebox(0,0)[rb]{\smash{{\SetFigFont{8}{9.6}{\familydefault}{\mddefault}{\updefault}{\color[rgb]{0,0,0}$\geq$}%
}}}}
\put(603,-246){\makebox(0,0)[b]{\smash{{\SetFigFont{9}{10.8}{\familydefault}{\mddefault}{\updefault}{\color[rgb]{0,0,0}$3$}%
}}}}
\put(206,377){\makebox(0,0)[b]{\smash{{\SetFigFont{9}{10.8}{\familydefault}{\mddefault}{\updefault}{\color[rgb]{0,0,0}$1$}%
}}}}
\put(997,377){\makebox(0,0)[b]{\smash{{\SetFigFont{9}{10.8}{\familydefault}{\mddefault}{\updefault}{\color[rgb]{0,0,0}$2$}%
}}}}
\put(603,734){\makebox(0,0)[b]{\smash{{\SetFigFont{8}{9.6}{\familydefault}{\mddefault}{\updefault}{\color[rgb]{0,0,0}$\geq$}%
}}}}
\end{picture}%

%% file: cyclepos3upbs.pdf_t
\begin{picture}(0,0)%
\includegraphics{cyclepos3upbs.pdf}%
\end{picture}%
\setlength{\unitlength}{3947sp}%
\begingroup\makeatletter\ifx\SetFigFont\undefined%
\gdef\SetFigFont#1#2#3#4#5{%
  \reset@font\fontsize{#1}{#2pt}%
  \fontfamily{#3}\fontseries{#4}\fontshape{#5}%
  \selectfont}%
\fi\endgroup%
\begin{picture}(1440,1184)(-142,-332)
\put(1018,-30){\makebox(0,0)[lb]{\smash{{\SetFigFont{8}{9.6}{\familydefault}{\mddefault}{\updefault}{\color[rgb]{0,0,0}$<$}%
}}}}
\put(178,-28){\makebox(0,0)[rb]{\smash{{\SetFigFont{8}{9.6}{\familydefault}{\mddefault}{\updefault}{\color[rgb]{0,0,0}$\geq$}%
}}}}
\put(603,-246){\makebox(0,0)[b]{\smash{{\SetFigFont{9}{10.8}{\familydefault}{\mddefault}{\updefault}{\color[rgb]{0,0,0}$3$}%
}}}}
\put(206,377){\makebox(0,0)[b]{\smash{{\SetFigFont{9}{10.8}{\familydefault}{\mddefault}{\updefault}{\color[rgb]{0,0,0}$1$}%
}}}}
\put(997,377){\makebox(0,0)[b]{\smash{{\SetFigFont{9}{10.8}{\familydefault}{\mddefault}{\updefault}{\color[rgb]{0,0,0}$2$}%
}}}}
\put(603,734){\makebox(0,0)[b]{\smash{{\SetFigFont{8}{9.6}{\familydefault}{\mddefault}{\updefault}{\color[rgb]{0,0,0}$\geq$}%
}}}}
\end{picture}%

%% file: cyclepos3upseq.pdf_t
\begin{picture}(0,0)%
\includegraphics{cyclepos3upseq.pdf}%
\end{picture}%
\setlength{\unitlength}{3947sp}%
\begingroup\makeatletter\ifx\SetFigFont\undefined%
\gdef\SetFigFont#1#2#3#4#5{%
  \reset@font\fontsize{#1}{#2pt}%
  \fontfamily{#3}\fontseries{#4}\fontshape{#5}%
  \selectfont}%
\fi\endgroup%
\begin{picture}(1440,1191)(-142,-339)
\put(1018,-30){\makebox(0,0)[lb]{\smash{{\SetFigFont{8}{9.6}{\familydefault}{\mddefault}{\updefault}{\color[rgb]{0,0,0}$<$}%
}}}}
\put(178,-28){\makebox(0,0)[rb]{\smash{{\SetFigFont{8}{9.6}{\familydefault}{\mddefault}{\updefault}{\color[rgb]{0,0,0}$\geq$}%
}}}}
\put(603,-246){\makebox(0,0)[b]{\smash{{\SetFigFont{9}{10.8}{\familydefault}{\mddefault}{\updefault}{\color[rgb]{0,0,0}$3$}%
}}}}
\put(206,377){\makebox(0,0)[b]{\smash{{\SetFigFont{9}{10.8}{\familydefault}{\mddefault}{\updefault}{\color[rgb]{0,0,0}$1$}%
}}}}
\put(997,377){\makebox(0,0)[b]{\smash{{\SetFigFont{9}{10.8}{\familydefault}{\mddefault}{\updefault}{\color[rgb]{0,0,0}$2$}%
}}}}
\put(603,734){\makebox(0,0)[b]{\smash{{\SetFigFont{8}{9.6}{\familydefault}{\mddefault}{\updefault}{\color[rgb]{0,0,0}$<$}%
}}}}
\end{picture}%

%% file: cyclepos3transpar.pdf_t
\begin{picture}(0,0)%
\includegraphics{cyclepos3transpar.pdf}%
\end{picture}%
\setlength{\unitlength}{3947sp}%
\begingroup\makeatletter\ifx\SetFigFont\undefined%
\gdef\SetFigFont#1#2#3#4#5{%
  \reset@font\fontsize{#1}{#2pt}%
  \fontfamily{#3}\fontseries{#4}\fontshape{#5}%
  \selectfont}%
\fi\endgroup%
\begin{picture}(1974,1267)(139,-416)
\put(1651,539){\makebox(0,0)[b]{\smash{{\SetFigFont{9}{10.8}{\familydefault}{\mddefault}{\updefault}{\color[rgb]{0,0,0}$111$}%
}}}}
\put(1351, 89){\makebox(0,0)[b]{\smash{{\SetFigFont{9}{10.8}{\familydefault}{\mddefault}{\updefault}{\color[rgb]{0,0,0}$011$}%
}}}}
\put(1951, 89){\makebox(0,0)[b]{\smash{{\SetFigFont{9}{10.8}{\familydefault}{\mddefault}{\updefault}{\color[rgb]{0,0,0}$101$}%
}}}}
\put(1651,-361){\makebox(0,0)[b]{\smash{{\SetFigFont{9}{10.8}{\familydefault}{\mddefault}{\updefault}{\color[rgb]{0,0,0}$110$}%
}}}}
\put(601,539){\makebox(0,0)[b]{\smash{{\SetFigFont{9}{10.8}{\familydefault}{\mddefault}{\updefault}{\color[rgb]{0,0,0}$000$}%
}}}}
\put(301, 89){\makebox(0,0)[b]{\smash{{\SetFigFont{9}{10.8}{\familydefault}{\mddefault}{\updefault}{\color[rgb]{0,0,0}$001$}%
}}}}
\put(901, 89){\makebox(0,0)[b]{\smash{{\SetFigFont{9}{10.8}{\familydefault}{\mddefault}{\updefault}{\color[rgb]{0,0,0}$100$}%
}}}}
\put(601,-361){\makebox(0,0)[b]{\smash{{\SetFigFont{9}{10.8}{\familydefault}{\mddefault}{\updefault}{\color[rgb]{0,0,0}$010$}%
}}}}
\end{picture}%

%% file: cyclepos3transbs.pdf_t
\begin{picture}(0,0)%
\includegraphics{cyclepos3transbs.pdf}%
\end{picture}%
\setlength{\unitlength}{3947sp}%
\begingroup\makeatletter\ifx\SetFigFont\undefined%
\gdef\SetFigFont#1#2#3#4#5{%
  \reset@font\fontsize{#1}{#2pt}%
  \fontfamily{#3}\fontseries{#4}\fontshape{#5}%
  \selectfont}%
\fi\endgroup%
\begin{picture}(1974,1224)(139,-373)
\put(376,389){\makebox(0,0)[b]{\smash{{\SetFigFont{9}{10.8}{\familydefault}{\mddefault}{\updefault}{\color[rgb]{0,0,0}$000$}%
}}}}
\put(901,389){\makebox(0,0)[b]{\smash{{\SetFigFont{9}{10.8}{\familydefault}{\mddefault}{\updefault}{\color[rgb]{0,0,0}$111$}%
}}}}
\put(1426,389){\makebox(0,0)[b]{\smash{{\SetFigFont{9}{10.8}{\familydefault}{\mddefault}{\updefault}{\color[rgb]{0,0,0}$100$}%
}}}}
\put(1876,389){\makebox(0,0)[b]{\smash{{\SetFigFont{9}{10.8}{\familydefault}{\mddefault}{\updefault}{\color[rgb]{0,0,0}$011$}%
}}}}
\put(376,-136){\makebox(0,0)[b]{\smash{{\SetFigFont{9}{10.8}{\familydefault}{\mddefault}{\updefault}{\color[rgb]{0,0,0}$010$}%
}}}}
\put(901,-136){\makebox(0,0)[b]{\smash{{\SetFigFont{9}{10.8}{\familydefault}{\mddefault}{\updefault}{\color[rgb]{0,0,0}$101$}%
}}}}
\put(1426,-136){\makebox(0,0)[b]{\smash{{\SetFigFont{9}{10.8}{\familydefault}{\mddefault}{\updefault}{\color[rgb]{0,0,0}$001$}%
}}}}
\put(1876,-136){\makebox(0,0)[b]{\smash{{\SetFigFont{9}{10.8}{\familydefault}{\mddefault}{\updefault}{\color[rgb]{0,0,0}$110$}%
}}}}
\end{picture}%

%% file: cyclepos3transseq.pdf_t
\begin{picture}(0,0)%
\includegraphics{cyclepos3transseq.pdf}%
\end{picture}%
\setlength{\unitlength}{3947sp}%
\begingroup\makeatletter\ifx\SetFigFont\undefined%
\gdef\SetFigFont#1#2#3#4#5{%
  \reset@font\fontsize{#1}{#2pt}%
  \fontfamily{#3}\fontseries{#4}\fontshape{#5}%
  \selectfont}%
\fi\endgroup%
\begin{picture}(1974,1224)(139,-373)
\put(601,389){\makebox(0,0)[b]{\smash{{\SetFigFont{9}{10.8}{\familydefault}{\mddefault}{\updefault}{\color[rgb]{0,0,0}$000$}%
}}}}
\put(1651,389){\makebox(0,0)[b]{\smash{{\SetFigFont{9}{10.8}{\familydefault}{\mddefault}{\updefault}{\color[rgb]{0,0,0}$111$}%
}}}}
\put(301, 14){\makebox(0,0)[b]{\smash{{\SetFigFont{9}{10.8}{\familydefault}{\mddefault}{\updefault}{\color[rgb]{0,0,0}$010$}%
}}}}
\put(601,-136){\makebox(0,0)[b]{\smash{{\SetFigFont{9}{10.8}{\familydefault}{\mddefault}{\updefault}{\color[rgb]{0,0,0}$100$}%
}}}}
\put(901, 14){\makebox(0,0)[b]{\smash{{\SetFigFont{9}{10.8}{\familydefault}{\mddefault}{\updefault}{\color[rgb]{0,0,0}$110$}%
}}}}
\put(1351, 14){\makebox(0,0)[b]{\smash{{\SetFigFont{9}{10.8}{\familydefault}{\mddefault}{\updefault}{\color[rgb]{0,0,0}$001$}%
}}}}
\put(1651,-136){\makebox(0,0)[b]{\smash{{\SetFigFont{9}{10.8}{\familydefault}{\mddefault}{\updefault}{\color[rgb]{0,0,0}$011$}%
}}}}
\put(1951, 14){\makebox(0,0)[b]{\smash{{\SetFigFont{9}{10.8}{\familydefault}{\mddefault}{\updefault}{\color[rgb]{0,0,0}$101$}%
}}}}
\end{picture}%

%% file: bp_diagram.pdf_t
\begin{picture}(0,0)%
\includegraphics{bp_diagram.pdf}%
\end{picture}%
\setlength{\unitlength}{3947sp}%
\begingroup\makeatletter\ifx\SetFigFont\undefined%
\gdef\SetFigFont#1#2#3#4#5{%
  \reset@font\fontsize{#1}{#2pt}%
  \fontfamily{#3}\fontseries{#4}\fontshape{#5}%
  \selectfont}%
\fi\endgroup%
\begin{picture}(5248,1224)(136,-373)
\put(451,-286){\makebox(0,0)[b]{\smash{{\SetFigFont{9}{10.8}{\rmdefault}{\mddefault}{\updefault}{\color[rgb]{0,0,0}$1$}%
}}}}
\put(151,-286){\makebox(0,0)[b]{\smash{{\SetFigFont{9}{10.8}{\rmdefault}{\mddefault}{\updefault}{\color[rgb]{0,0,0}$0$}%
}}}}
\put(151,314){\makebox(0,0)[b]{\smash{{\SetFigFont{9}{10.8}{\rmdefault}{\mddefault}{\updefault}{\color[rgb]{0,0,0}$2$}%
}}}}
\put(151,539){\makebox(0,0)[b]{\smash{{\SetFigFont{9}{10.8}{\rmdefault}{\mddefault}{\updefault}{\color[rgb]{0,0,0}$1$}%
}}}}
\put(451,539){\makebox(0,0)[b]{\smash{{\SetFigFont{9}{10.8}{\rmdefault}{\mddefault}{\updefault}{\color[rgb]{0,0,0}$1$}%
}}}}
\put(751,539){\makebox(0,0)[b]{\smash{{\SetFigFont{9}{10.8}{\rmdefault}{\mddefault}{\updefault}{\color[rgb]{0,0,0}$1$}%
}}}}
\put(1051,539){\makebox(0,0)[b]{\smash{{\SetFigFont{9}{10.8}{\rmdefault}{\mddefault}{\updefault}{\color[rgb]{0,0,0}$1$}%
}}}}
\put(1351,539){\makebox(0,0)[b]{\smash{{\SetFigFont{9}{10.8}{\rmdefault}{\mddefault}{\updefault}{\color[rgb]{0,0,0}$1$}%
}}}}
\put(1651,539){\makebox(0,0)[b]{\smash{{\SetFigFont{9}{10.8}{\rmdefault}{\mddefault}{\updefault}{\color[rgb]{0,0,0}$1$}%
}}}}
\put(1951,539){\makebox(0,0)[b]{\smash{{\SetFigFont{9}{10.8}{\rmdefault}{\mddefault}{\updefault}{\color[rgb]{0,0,0}$1$}%
}}}}
\put(2251,539){\makebox(0,0)[b]{\smash{{\SetFigFont{9}{10.8}{\rmdefault}{\mddefault}{\updefault}{\color[rgb]{0,0,0}$1$}%
}}}}
\put(2551,539){\makebox(0,0)[b]{\smash{{\SetFigFont{9}{10.8}{\rmdefault}{\mddefault}{\updefault}{\color[rgb]{0,0,0}$1$}%
}}}}
\put(451,314){\makebox(0,0)[b]{\smash{{\SetFigFont{9}{10.8}{\rmdefault}{\mddefault}{\updefault}{\color[rgb]{0,0,0}$3$}%
}}}}
\put(751,314){\makebox(0,0)[b]{\smash{{\SetFigFont{9}{10.8}{\rmdefault}{\mddefault}{\updefault}{\color[rgb]{0,0,0}$2$}%
}}}}
\put(1351,314){\makebox(0,0)[b]{\smash{{\SetFigFont{9}{10.8}{\rmdefault}{\mddefault}{\updefault}{\color[rgb]{0,0,0}$2$}%
}}}}
\put(1951,314){\makebox(0,0)[b]{\smash{{\SetFigFont{9}{10.8}{\rmdefault}{\mddefault}{\updefault}{\color[rgb]{0,0,0}$2$}%
}}}}
\put(2551,314){\makebox(0,0)[b]{\smash{{\SetFigFont{9}{10.8}{\rmdefault}{\mddefault}{\updefault}{\color[rgb]{0,0,0}$2$}%
}}}}
\put(1051,314){\makebox(0,0)[b]{\smash{{\SetFigFont{9}{10.8}{\rmdefault}{\mddefault}{\updefault}{\color[rgb]{0,0,0}$3$}%
}}}}
\put(1651,314){\makebox(0,0)[b]{\smash{{\SetFigFont{9}{10.8}{\rmdefault}{\mddefault}{\updefault}{\color[rgb]{0,0,0}$3$}%
}}}}
\put(2251,314){\makebox(0,0)[b]{\smash{{\SetFigFont{9}{10.8}{\rmdefault}{\mddefault}{\updefault}{\color[rgb]{0,0,0}$3$}%
}}}}
\put(451, 89){\makebox(0,0)[b]{\smash{{\SetFigFont{9}{10.8}{\rmdefault}{\mddefault}{\updefault}{\color[rgb]{0,0,0}$5$}%
}}}}
\put(751, 89){\makebox(0,0)[b]{\smash{{\SetFigFont{9}{10.8}{\rmdefault}{\mddefault}{\updefault}{\color[rgb]{0,0,0}$6$}%
}}}}
\put(151, 89){\makebox(0,0)[b]{\smash{{\SetFigFont{9}{10.8}{\rmdefault}{\mddefault}{\updefault}{\color[rgb]{0,0,0}$4$}%
}}}}
\put(1051, 89){\makebox(0,0)[b]{\smash{{\SetFigFont{9}{10.8}{\rmdefault}{\mddefault}{\updefault}{\color[rgb]{0,0,0}$4$}%
}}}}
\put(1951, 89){\makebox(0,0)[b]{\smash{{\SetFigFont{9}{10.8}{\rmdefault}{\mddefault}{\updefault}{\color[rgb]{0,0,0}$4$}%
}}}}
\put(1351, 89){\makebox(0,0)[b]{\smash{{\SetFigFont{9}{10.8}{\rmdefault}{\mddefault}{\updefault}{\color[rgb]{0,0,0}$5$}%
}}}}
\put(2251, 89){\makebox(0,0)[b]{\smash{{\SetFigFont{9}{10.8}{\rmdefault}{\mddefault}{\updefault}{\color[rgb]{0,0,0}$5$}%
}}}}
\put(1651, 89){\makebox(0,0)[b]{\smash{{\SetFigFont{9}{10.8}{\rmdefault}{\mddefault}{\updefault}{\color[rgb]{0,0,0}$6$}%
}}}}
\put(2551, 89){\makebox(0,0)[b]{\smash{{\SetFigFont{9}{10.8}{\rmdefault}{\mddefault}{\updefault}{\color[rgb]{0,0,0}$6$}%
}}}}
\put(1651,-286){\makebox(0,0)[b]{\smash{{\SetFigFont{9}{10.8}{\rmdefault}{\mddefault}{\updefault}{\color[rgb]{0,0,0}$5$}%
}}}}
\put(2926,-101){\makebox(0,0)[b]{\smash{{\SetFigFont{9}{10.8}{\rmdefault}{\mddefault}{\updefault}{\color[rgb]{0,0,0}$t$}%
}}}}
\put(3751,489){\makebox(0,0)[b]{\smash{{\SetFigFont{9}{10.8}{\rmdefault}{\mddefault}{\updefault}{\color[rgb]{0,0,0}$1$}%
}}}}
\put(4501,489){\makebox(0,0)[b]{\smash{{\SetFigFont{9}{10.8}{\rmdefault}{\mddefault}{\updefault}{\color[rgb]{0,0,0}$2$}%
}}}}
\put(5251,489){\makebox(0,0)[b]{\smash{{\SetFigFont{9}{10.8}{\rmdefault}{\mddefault}{\updefault}{\color[rgb]{0,0,0}$3$}%
}}}}
\put(3751,-111){\makebox(0,0)[b]{\smash{{\SetFigFont{9}{10.8}{\rmdefault}{\mddefault}{\updefault}{\color[rgb]{0,0,0}$4$}%
}}}}
\put(4501,-111){\makebox(0,0)[b]{\smash{{\SetFigFont{9}{10.8}{\rmdefault}{\mddefault}{\updefault}{\color[rgb]{0,0,0}$5$}%
}}}}
\put(5251,-111){\makebox(0,0)[b]{\smash{{\SetFigFont{9}{10.8}{\rmdefault}{\mddefault}{\updefault}{\color[rgb]{0,0,0}$6$}%
}}}}
\end{picture}%

%% file: netplant.pdf_t
\begin{picture}(0,0)%
\includegraphics{netplant.pdf}%
\end{picture}%
\setlength{\unitlength}{3947sp}%
\begingroup\makeatletter\ifx\SetFigFont\undefined%
\gdef\SetFigFont#1#2#3#4#5{%
  \reset@font\fontsize{#1}{#2pt}%
  \fontfamily{#3}\fontseries{#4}\fontshape{#5}%
  \selectfont}%
\fi\endgroup%
\begin{picture}(2210,2103)(-40,-1251)
\put(853,-29){\makebox(0,0)[rb]{\smash{{\SetFigFont{8}{9.6}{\familydefault}{\mddefault}{\updefault}{\color[rgb]{0,0,0}$-2$}%
}}}}
\put(1278,733){\makebox(0,0)[b]{\smash{{\SetFigFont{8}{9.6}{\familydefault}{\mddefault}{\updefault}{\color[rgb]{0,0,0}$-2$}%
}}}}
\put(1693,-31){\makebox(0,0)[lb]{\smash{{\SetFigFont{8}{9.6}{\familydefault}{\mddefault}{\updefault}{\color[rgb]{0,0,0}$-2$}%
}}}}
\put(1278,478){\makebox(0,0)[b]{\smash{{\SetFigFont{8}{9.6}{\familydefault}{\mddefault}{\updefault}{\color[rgb]{0,0,0}$-2$}%
}}}}
\put(1279,-598){\makebox(0,0)[b]{\smash{{\SetFigFont{8}{9.6}{\familydefault}{\mddefault}{\updefault}{\color[rgb]{0,0,0}$+1$}%
}}}}
\put(881,-1036){\makebox(0,0)[b]{\smash{{\SetFigFont{9}{10.8}{\familydefault}{\mddefault}{\updefault}{\color[rgb]{0,0,0}$4$}%
}}}}
\put(1670,-1039){\makebox(0,0)[b]{\smash{{\SetFigFont{9}{10.8}{\familydefault}{\mddefault}{\updefault}{\color[rgb]{0,0,0}$5$}%
}}}}
\put(578,-1026){\makebox(0,0)[rb]{\smash{{\SetFigFont{8}{9.6}{\familydefault}{\mddefault}{\updefault}{\color[rgb]{0,0,0}$+1$}%
}}}}
\put(1279,-1205){\makebox(0,0)[b]{\smash{{\SetFigFont{8}{9.6}{\familydefault}{\mddefault}{\updefault}{\color[rgb]{0,0,0}$+1$}%
}}}}
\put(448,-340){\makebox(0,0)[rb]{\smash{{\SetFigFont{8}{9.6}{\familydefault}{\mddefault}{\updefault}{\color[rgb]{0,0,0}$-2$}%
}}}}
\put(916,-513){\makebox(0,0)[rb]{\smash{{\SetFigFont{8}{9.6}{\familydefault}{\mddefault}{\updefault}{\color[rgb]{0,0,0}$-2$}%
}}}}
\put(1861,-518){\makebox(0,0)[lb]{\smash{{\SetFigFont{8}{9.6}{\familydefault}{\mddefault}{\updefault}{\color[rgb]{0,0,0}$-2$}%
}}}}
\put(1273,-858){\makebox(0,0)[b]{\smash{{\SetFigFont{8}{9.6}{\familydefault}{\mddefault}{\updefault}{\color[rgb]{0,0,0}$-2$}%
}}}}
\put(1584, 53){\makebox(0,0)[rb]{\smash{{\SetFigFont{8}{9.6}{\familydefault}{\mddefault}{\updefault}{\color[rgb]{0,0,0}$-2$}%
}}}}
\put(969, 55){\makebox(0,0)[lb]{\smash{{\SetFigFont{8}{9.6}{\familydefault}{\mddefault}{\updefault}{\color[rgb]{0,0,0}$-2$}%
}}}}
\put(623,537){\makebox(0,0)[rb]{\smash{{\SetFigFont{8}{9.6}{\familydefault}{\mddefault}{\updefault}{\color[rgb]{0,0,0}$+1$}%
}}}}
\put(1913,535){\makebox(0,0)[lb]{\smash{{\SetFigFont{8}{9.6}{\familydefault}{\mddefault}{\updefault}{\color[rgb]{0,0,0}$+1$}%
}}}}
\put(320,692){\makebox(0,0)[rb]{\smash{{\SetFigFont{9}{10.8}{\familydefault}{\mddefault}{\updefault}{\color[rgb]{0,0,0}$G(f):$}%
}}}}
\put(1278,-247){\makebox(0,0)[b]{\smash{{\SetFigFont{9}{10.8}{\familydefault}{\mddefault}{\updefault}{\color[rgb]{0,0,0}$3$}%
}}}}
\put(881,376){\makebox(0,0)[b]{\smash{{\SetFigFont{9}{10.8}{\familydefault}{\mddefault}{\updefault}{\color[rgb]{0,0,0}$1$}%
}}}}
\put(1672,376){\makebox(0,0)[b]{\smash{{\SetFigFont{9}{10.8}{\familydefault}{\mddefault}{\updefault}{\color[rgb]{0,0,0}$2$}%
}}}}
\end{picture}%

%% file: bp_diagram_plant.pdf_t
\begin{picture}(0,0)%
\includegraphics{bp_diagram_plant.pdf}%
\end{picture}%
\setlength{\unitlength}{3947sp}%
\begingroup\makeatletter\ifx\SetFigFont\undefined%
\gdef\SetFigFont#1#2#3#4#5{%
  \reset@font\fontsize{#1}{#2pt}%
  \fontfamily{#3}\fontseries{#4}\fontshape{#5}%
  \selectfont}%
\fi\endgroup%
\begin{picture}(3365,266)(168,406)
\put(2551,489){\makebox(0,0)[b]{\smash{{\SetFigFont{9}{10.8}{\rmdefault}{\mddefault}{\updefault}{\color[rgb]{0,0,0}$4$}%
}}}}
\put(3301,489){\makebox(0,0)[b]{\smash{{\SetFigFont{9}{10.8}{\rmdefault}{\mddefault}{\updefault}{\color[rgb]{0,0,0}$5$}%
}}}}
\put(301,489){\makebox(0,0)[b]{\smash{{\SetFigFont{9}{10.8}{\rmdefault}{\mddefault}{\updefault}{\color[rgb]{0,0,0}$1$}%
}}}}
\put(1051,489){\makebox(0,0)[b]{\smash{{\SetFigFont{9}{10.8}{\rmdefault}{\mddefault}{\updefault}{\color[rgb]{0,0,0}$2$}%
}}}}
\put(1801,489){\makebox(0,0)[b]{\smash{{\SetFigFont{9}{10.8}{\rmdefault}{\mddefault}{\updefault}{\color[rgb]{0,0,0}$3$}%
}}}}
\end{picture}%

%% file: cardionet.pdf_t
\begin{picture}(0,0)%
\includegraphics{cardionet.pdf}%
\end{picture}%
\setlength{\unitlength}{3947sp}%
\begingroup\makeatletter\ifx\SetFigFont\undefined%
\gdef\SetFigFont#1#2#3#4#5{%
  \reset@font\fontsize{#1}{#2pt}%
  \fontfamily{#3}\fontseries{#4}\fontshape{#5}%
  \selectfont}%
\fi\endgroup%
\begin{picture}(2003,1532)(-40,-680)
\put(1473,-378){\makebox(0,0)[b]{\smash{{\SetFigFont{9}{10.8}{\familydefault}{\mddefault}{\updefault}{\color[rgb]{0,0,0}$3$}%
}}}}
\put(  1,689){\makebox(0,0)[lb]{\smash{{\SetFigFont{9}{10.8}{\familydefault}{\mddefault}{\updefault}{\color[rgb]{0,0,0}$G(g):$}%
}}}}
\put(1754,453){\makebox(0,0)[lb]{\smash{{\SetFigFont{8}{9.6}{\familydefault}{\mddefault}{\updefault}{\color[rgb]{0,0,0}$+1$}%
}}}}
\put(1472,458){\makebox(0,0)[b]{\smash{{\SetFigFont{9}{10.8}{\familydefault}{\mddefault}{\updefault}{\color[rgb]{0,0,0}$2$}%
}}}}
\put(560, 53){\makebox(0,0)[lb]{\smash{{\SetFigFont{8}{9.6}{\familydefault}{\mddefault}{\updefault}{\color[rgb]{0,0,0}$+1$}%
}}}}
\put(1537, 53){\makebox(0,0)[rb]{\smash{{\SetFigFont{8}{9.6}{\familydefault}{\mddefault}{\updefault}{\color[rgb]{0,0,0}$+1$}%
}}}}
\put(634,-375){\makebox(0,0)[b]{\smash{{\SetFigFont{9}{10.8}{\familydefault}{\mddefault}{\updefault}{\color[rgb]{0,0,0}$4$}%
}}}}
\put(631,457){\makebox(0,0)[b]{\smash{{\SetFigFont{9}{10.8}{\familydefault}{\mddefault}{\updefault}{\color[rgb]{0,0,0}$1$}%
}}}}
\put(409, 53){\makebox(0,0)[rb]{\smash{{\SetFigFont{8}{9.6}{\familydefault}{\mddefault}{\updefault}{\color[rgb]{0,0,0}$-1$}%
}}}}
\put(1052,716){\makebox(0,0)[b]{\smash{{\SetFigFont{8}{9.6}{\familydefault}{\mddefault}{\updefault}{\color[rgb]{0,0,0}$-2$}%
}}}}
\put(1051,499){\makebox(0,0)[b]{\smash{{\SetFigFont{8}{9.6}{\familydefault}{\mddefault}{\updefault}{\color[rgb]{0,0,0}$+2$}%
}}}}
\put(1693, 50){\makebox(0,0)[lb]{\smash{{\SetFigFont{8}{9.6}{\familydefault}{\mddefault}{\updefault}{\color[rgb]{0,0,0}$-1$}%
}}}}
\put(1048,-404){\makebox(0,0)[b]{\smash{{\SetFigFont{8}{9.6}{\familydefault}{\mddefault}{\updefault}{\color[rgb]{0,0,0}$+1$}%
}}}}
\put(1051,-634){\makebox(0,0)[b]{\smash{{\SetFigFont{8}{9.6}{\familydefault}{\mddefault}{\updefault}{\color[rgb]{0,0,0}$-1$}%
}}}}
\put(348,-353){\makebox(0,0)[rb]{\smash{{\SetFigFont{8}{9.6}{\familydefault}{\mddefault}{\updefault}{\color[rgb]{0,0,0}$+1$}%
}}}}
\end{picture}%